\documentclass[twocolumn,astrosymb]{aastex631}

\newcommand{\psr}{PSR~J1032$-$5804}
\newcommand{\vast}{VAST~J103250.4--580434}
\newcommand{\change}[1]{#1}
\newcommand{\Change}[1]{#1}

\received{XXX}
\revised{YYY}
\accepted{ZZZ}

\submitjournal{ApJ}

\shorttitle{Discovery of \psr}
\shortauthors{Wang et al.}

\begin{document}

\title{Discovery of a young, highly scattered pulsar \psr\ with the Australian SKA Pathfinder}

\correspondingauthor{Ziteng Wang}
\email{ziteng.wang@curtin.edu.au}

\author[0000-0002-2066-9823]{Ziteng Wang}
\affiliation{International Centre for Radio Astronomy Research, Curtin University, Bentley, WA 6102, Australia}

\author[0000-0001-6295-2881]{David~L.~Kaplan}
\affiliation{Center for Gravitation, Cosmology, and Astrophysics, Department of Physics, University of Wisconsin-Milwaukee, P.O. Box 413, Milwaukee, WI 53201, USA}

\author[0000-0002-9409-3214]{Rahul Sengar}
\affiliation{Center for Gravitation, Cosmology, and Astrophysics, Department of Physics, University of Wisconsin-Milwaukee, P.O. Box 413, Milwaukee, WI 53201, USA}

\author[0000-0002-9994-1593]{Emil Lenc}
\affiliation{ATNF, CSIRO Space and Astronomy, PO Box 76, Epping, New South Wales 1710, Australia}

\author[0000-0002-9583-2947]{Andrew Zic}
\affiliation{ATNF, CSIRO Space and Astronomy, PO Box 76, Epping, New South Wales 1710, Australia}

\author[0000-0002-8935-9882]{Akash Anumarlapudi}
\affiliation{Center for Gravitation, Cosmology, and Astrophysics, Department of Physics, University of Wisconsin-Milwaukee, P.O. Box 413, Milwaukee, WI 53201, USA}

\author[0000-0002-3382-9558]{B. M. Gaensler}
\affiliation{Dunlap Institute for Astronomy and Astrophysics, University of Toronto, 50 St. George St., Toronto, ON M5S 3H4, Canada}
\affiliation{David A. Dunlap Department of Astronomy and Astrophysics, University of Toronto, 50 St. George St., Toronto, ON M5S 3H4, Canada}
\affiliation{Division of Physical and Biological Sciences, University of California Santa Cruz, 1156 High Street, Santa Cruz, CA 95064, USA}

\author[0000-0002-5119-4808]{Natasha Hurley-Walker}
\affiliation{International Centre for Radio Astronomy Research, Curtin University, Bentley, WA 6102, Australia}

\author[0000-0002-2686-438X]{Tara Murphy}
\affiliation{Sydney Institute for Astronomy, School of Physics, University of Sydney, Sydney, New South Wales 2006, Australia.}
\affiliation{ARC Centre of Excellence for Gravitational Wave Discovery (OzGrav), Hawthorn, Victoria, Australia}

\author[0000-0003-0203-1196]{Yuanming Wang}
\affiliation{Centre for Astrophysics and Supercomputing, Swinburne University of Technology, Hawthorn, VIC 3122, Australia}
\affiliation{ARC Centre of Excellence for Gravitational Wave Discovery (OzGrav), Hawthorn, Victoria, Australia}

\begin{abstract}

We report the discovery of a young, highly scattered pulsar in a search for highly circularly polarized radio sources as part of the Australian Square Kilometre Array Pathfinder (ASKAP) Variables and Slow Transients (VAST) survey.
In follow-up observations with Murriyang/Parkes, we identified \psr\ and measured  a period of 78.7\,ms, dispersion measure (DM) of \change{$819\pm4$}\,pc\,cm$^{-3}$, rotation measure of \change{$-2000\pm1\,$}rad\,m$^{-2}$, and a characteristic age of 34.6\,kyr.
We found a pulse scattering timescale at 3\,GHz of $\sim$22\,ms, implying a timescale at 1\,GHz of $\sim$3845\,ms, which is the third most scattered pulsar known and explains its non-detection in previous pulsar surveys.
We discuss the identification of a possible pulsar wind nebula and supernova remnant in the pulsar's local environment by analyzing the pulsar spectral energy distribution and the surrounding extended emission from multiwavelength  images. 
Our result highlights the possibility of identifying extremely scattered pulsars from radio continuum images. 
Ongoing and future large-scale radio continuum surveys will offer us an unprecedented opportunity to find more extreme pulsars (e.g., highly scattered, highly intermittent, highly accelerated), which will enhance our understanding of the characteristics of pulsars and the interstellar medium.

\end{abstract}




\section{Introduction} \label{sec:intro}


When radio pulses from pulsars traverse the turbulent interstellar medium, multi-path propagation leads to temporal and spatial scattering, which in turn smears out the pulse profile.
Despite extensive pulsar searching surveys spanning decades \change{\citep[e.g.,][]{1998MNRAS.295..743L, 2001MNRAS.328...17M, 2004hpa..book.....L, 2010MNRAS.409..619K, 2014ApJ...791...67S, 2018MNRAS.473..116K}}, some highly scattered pulsars remain challenging to detect with traditional time-domain techniques. 
These pulsars are especially hard to detect at frequencies $\lesssim 1\,$GHz where the majority of pulsar surveys take place.
The periodic signal is difficult (or impossible) to detect when the observed pulse profile width is comparable to the pulsar spin period.
For a Kolmogorov distribution of interstellar medium inhomogeneities, the scattering timescale $\tau$ scales with frequency $\nu$ as  $\tau \propto \nu^{-4.4}$  \citep[e.g.,][]{1986MNRAS.220...19R}.
Most of the pulsar searching surveys are not sensitive to highly scattered pulsars as they are performed at a relatively low frequency, such as the High Time Resolution Universe Pulsar Survey \citep{2010MNRAS.409..619K} at $\sim$1.4\,GHz, the Green Bank Northern Celestial Cap Pulsar Survey \citep{2014ApJ...791...67S} at 350\,MHz, and the MPIfR-MeerKAT Galactic Plane survey \citep[MMGPS;][]{2023MNRAS.524.1291P} currently at $\sim1.4\,$GHz \change{(from 856\,MHz to 1712\,MHz)}.

As a concrete example, the most scattered pulsar found to date, \object[PSR J1813-1749]{PSR~J1813$-$1749}, was first identified as a pulsar candidate as a  TeV source \citep{2005Sci...307.1938A, 2006ApJ...636..777A}, an X-ray source \citep{2005ApJ...629L.105B, 2005ApJ...629L.109U}, and a supernova remnant association \citep{2005ApJ...629L.105B}.
\citet{2009ApJ...700L.158G} measured the pulse period $P=44.7\,$ms in X-rays, while no radio pulsations were detected with the Green Bank Telescope (GBT) at 1.4\,GHz and 2\,GHz \citep{2012ApJ...753L..14H} or the Effelsberg Telescope at 1.4\,GHz \citep{2018ApJ...866..100D}, though radio interferometric observations did detect a variable point source at the pulsar position \citep{2010RMxAA..46..153D, 2018ApJ...866..100D}.
Radio pulsations were finally detected with the GBT at frequencies of 4.4--10.2\,GHz \citep{2021ApJ...917...67C}. The pulsar was detected with a high scattering timescale $\tau \approx 0.25\,$s at 2\,GHz, which explains the non-detection in the previous pulsar searches.

The special properties of pulsars can help in identifying pulsar candidates in radio continuum images \citep[e.g.,][]{2023PASA...40....3S}.
Pulsars typically have steeper spectra than most other radio source types, with spectral indices $\alpha < -1$, where $S_\nu \approx \nu^\alpha$ \change{\citep[e.g.,][]{2013MNRAS.431.1352B, 2023MNRAS.520.4582P}}.
Selecting steep-spectrum radio sources has long been used to find new pulsar candidates, and has successfully identified new pulsars including the first detected millisecond pulsar \citep{1982Natur.300..615B}.

Pulsars are also one of a few astronomical sources measured to be strongly polarized \citep[e.g.,][]{2004hpa..book.....L, 2018MNRAS.474.4629J}, and in particular circularly polarized. This means searches in circular polarisation are another method that can be used to find them in the image domain \citep[e.g.,][]{1998MNRAS.296..813G}. For example, 
\citet{2018MNRAS.478.2835L} conducted an all-sky circular polarization survey at 200\,MHz with the Murchison Widefield Array \citep[MWA;][]{2013PASA...30...31B} and identified 33 known pulsars. \citet{2021MNRAS.502.5438P} performed a circular polarization survey with the Australian Square Kilometre Array Pathfinder \citep[ASKAP;][]{2008ExA....22..151J, 2021PASA...38....9H} on the Rapid ASKAP Continuum Survey \citep[RACS;][]{2020PASA...37...48M} data and identified 37 known pulsars. In a similar search, \citet{2019ApJ...884...96K} serendipitously discovered a new millisecond pulsar, PSR~J1431--6328, by identifying circularly polarized sources with the ASKAP data.
Recently, \citet{2022A&A...661A..87S} discovered two new pulsars, PSR~J1049+5822 and PSR~J1602+3901, with the Low-Frequency Array (LOFAR) Two-metre Sky Survey (LoTSS) as part of the Targeted search using LoTSS images for polarized pulsars (TULIPP) survey.

Finally, pulsars can show flux density variability due to, for example, pulse nulling \citep[e.g.,][]{1970Natur.228...42B}, pulse intermittency \citep[e.g.,][]{2006Sci...312..549K}, interstellar scintillation \citep[e.g.,][]{1970MNRAS.150...67R}, and/or eclipsing by the companion \citep[e.g.,][]{2016MNRAS.459.2681B}.
Finding highly variable sources is another way to select pulsar candidates, but additional criteria (e.g., radio spectrum, polarization, multi-wavelength counterparts association) are often applied to make the sample size manageable. For example, \citet{2022ApJ...930...38W} identified 27 highly variable point sources towards the Magellanic Clouds with ASKAP, including a new pulsar. The new pulsar was the only source that was circularly polarized but which did not have a multi-wavelength counterpart.  \citet{2016MNRAS.462.3115D} discussed the prospect of identifying pulsars in variance images.
These image domain techniques can help us discover pulsars located in previously poorly explored parameter spaces, such as pulsars with high dispersion measure (DM), extreme nulling behaviors, and/or highly scattered pulses.

In this paper, we present the discovery of a young, highly scattered pulsar with the circular polarization search technique based on the data from two ASKAP  projects: the Variable and Slow Transients \citep[VAST;][]{2013PASA...30....6M, 2021PASA...38...54M} survey and the Evolutionary Map of the Universe \citep[EMU;][]{2011PASA...28..215N,2021PASA...38...46N} survey. We confirmed the nature of the source in a dedicated search with the Parkes radio telescope, Murriyang. In Section~\ref{sec:discover}, we summarise the discovery observations for the new pulsar. Analysis of archival observations and follow-up observations are discussed in Section~\ref{sec:follow}.
In Section~\ref{sec:discuss}, we discuss the properties and the local environment of the pulsar. We also discuss the prospects for identifying new pulsars, especially highly scattered ones, through image domain techniques.
Conclusions of this work are presented in Section~\ref{sec:concls}


\section{Source Discovery}\label{sec:discover}

\subsection{ASKAP Discovery}

As part of the VAST survey, we have been conducting ASKAP observations of the southern Galactic plane. \change{VAST observed 41 Galactic fields covering $|b| < 6\degr$ and $\delta < -10\degr$ totaling $1260\,{\rm deg}^2$, repeating each field, on average, every two weeks \footnote{\change{ASKAP uses dynamic scheduling so we cannot give an exact cadence}}} since 2022 November.
Each observation had  $\sim$12\,mins integration at a central frequency of 888\,MHz with a bandwidth of 288\,MHz, achieving a typical sensitivity of 0.4\,mJy\,beam$^{-1}$ for the fields covering the Galactic plane.
All four instrumental polarization products (XX, XY, YX, and YY) were recorded to allow images to be made in four Stokes parameters (I, Q, U, and V).
The data were processed \change{offline} using the \textsc{ASKAPSoft} pipeline \citep{cornwell2011askap}, from which we can get Stokes I/V images and catalogs.
\change{\object[PKS B1934-638]{PKS~B1934--638} was used for the flux scale and bandpass calibration, and self-calibration was applied to correct for phase variations during the observation.}
A detailed description of data reduction is given by \citet{2021PASA...38...54M}.

We conducted a search for highly circularly polarized sources in the VAST Galactic plane observations to identify interesting sources \citep[also see][for the results of other circular polarization searches with ASKAP data]{2021MNRAS.502.5438P,2023ApJ...951L..43R}.
Most of the detected sources were matched to known pulsars or  stellar objects. 
We selected \vast\ for further investigation because no clear multi-wavelength association was found.
\vast\ was first detected in a VAST observation on 2022 November 19 (Schedule Block 45739) with a Stokes I peak flux density of $4.16\pm0.52$\,mJy\,beam$^{-1}$, and a fractional circular polarization of $-28.5\pm5.0\%$: a higher magnitude than average for pulsars, but consistent with the population \citep{2023ApJ...956...28A}. 
The broader search results will be presented in future work.

\subsection{Murriyang Discovery}

Motivated by the detection of circular polarization and lack of multiwavelength association, we conducted a follow-up observation of \vast\ with the 64-m Parkes telescope, Murriyang, on 2023 February 24 (project code PX103) with the Ultra-Wideband Low (UWL) receiver \citep{2020PASA...37...12H}, which provides a wide frequency coverage spanning 704 to 4032\,MHz and has been pivotal in confirming pulsar candidates that are scarcely detectable below 1.4 GHz \citep{2023MNRAS.522.1071S}. 
The data were recorded  in 1-bit pulsar searching mode \change{(total intensity)} with the MEDUSA backend.
The observation was 2500\,seconds with a 32\,$\mu$s time resolution and 62.5\,kHz frequency resolution (2048 channels per 128\,MHz subband).

The periodic search was carried out with \textsc{pulsar\_miner} \footnote{\url{https://github.com/alex88ridolfi/PULSAR_MINER}}\citep{2021MNRAS.504.1407R}, an automated pulsar searching pipeline based on \textsc{presto}\footnote{\url{https://github.com/scottransom/presto}} \citep{2001PhDT.......123R}.
To run the search more efficiently, we divided the data into two segments (1250\,s each) and only selected two groups of subbands (bottom band group from 832 to 1216\,MHz \change{considering the steep spectral indices for the normal pulsar population}, and high band group from 2624 to 3008\,MHz \change{considering the potential high scattering and/or dispersion}).
The dispersion measure (DM) range that we searched was 2--1500\,pc\,cm$^{-3}$ \change{with \Change{harmonic summing of 8 and }no acceleration search applied}. \change{We only folded the candidates with a signal to noise ratio threshold of 8}.
We identified a strong pulsar candidate, now called \psr, with a period of 78.72\,ms and a DM of 867.8\,pc\,cm$^{-3}$ in the high band group data.  There is no previously published pulsar at this position in the ATNF pulsar catalog \citep{2005AJ....129.1993M}, nor is there any unpublished discovery in the \texttt{Pulsar Survey Scraper}\footnote{\url{https://pulsar.cgca-hub.org}} \citep{2022ascl.soft10001K}.

\psr\ was measured to have a wide pulse profile, with a duty cycle $\sim$30\% at $\sim$2.8\,GHz (using the pulse width at 50\% maximum).
After the initial discovery, we also folded the data at frequencies ranging from 1216 to 1600\,MHz (centered at 1408\,MHz) and from 3648 to 4032\,MHz (centered at 3840\,MHz) respectively. We saw a clear periodic signal (25.5$\sigma$) from the data centered at $\sim$3.8\,GHz, but no signal from the data centered at $\sim$1.4\,GHz.

\section{Follow-up Observations and Source Properties}\label{sec:follow}

\subsection{Timing and Polarization}

All timing observations were conducted at the Murriyang radio telescope using the UWL receiver and MEDUSA backend. Each observation was recorded in both fold and search mode with full Stokes parameters along with a noise diode observation. \change{All data were coherently de-dispersed at the best DM value we could measure at the time of the observation ($821\,$pc\,cm$^{-3}$)}. The time resolution for each observation was 128\,$\mu \rm s$, with a bandwidth of 0.5\,MHz per channel. Prior to the timing analysis, observations were first cleaned using \texttt{clfd}\footnote{\url{https://github.com/v-morello/clfd}}. We also excised each observation affected by narrowband and impulsive radio-frequency interference (RFI) using  \texttt{pazi} of the \texttt{PSRCHIVE} package. \change{Using the highest SNR observation we created a standard profile using \texttt{pas} from the \texttt{PSRCHIVE} package \citep{2004PASA...21..302H}. Note that the pulsar is not visible below 2.5\,GHz, therefore the UWL frequency band from 704--2500\,MHz was removed from the analysis. All observations of the pulsar taken on 11 epochs were frequency scrunched to one sub-band and polarization scrunched to total intensity for timing (polarization analysis was performed separately). We  divided the data into 1 to 4 sub-integrations depending on their signal to noise ratio. To generate the TOAs we used \texttt{pat} from \texttt{PSRCHIVE}, and employed the Fourier domain with Markov Chain Monte Carlo (FDM) algorithm. In total  26 TOAs were obtained over 11 epochs, and each TOA had signal to noise ratio greater than 15.}

\begin{figure}
    \includegraphics[width=0.95\columnwidth]{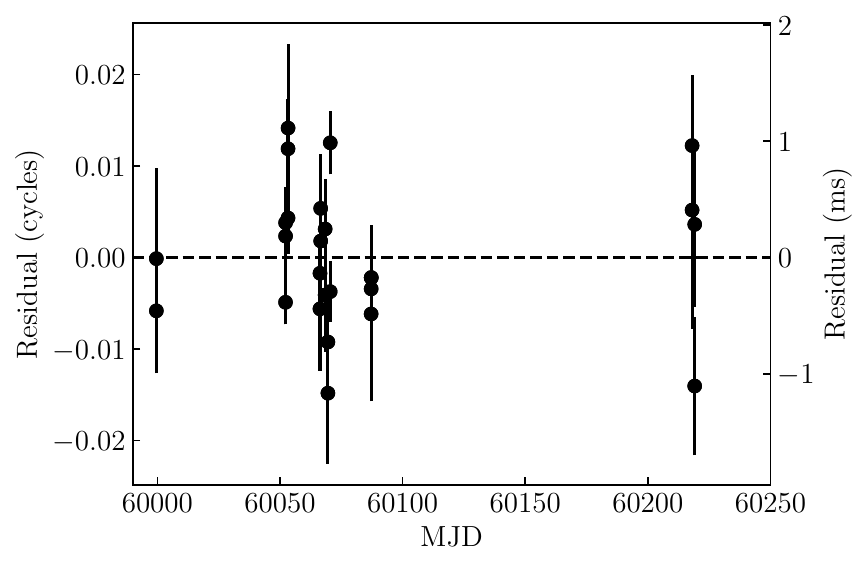}
    \caption{Timing residuals for \psr\ from Parkes/Murriyang data.  We plot the phase residuals (left y-axis) or time residuals (right y-axis) against the observation date (MJD).}
    \label{fig:resid}
\end{figure}

We fit the TOAs derived from the Parkes/Murriyang data in \texttt{PINT} \citep{2021ApJ...911...45L}, using only frequencies above 2.5\,GHz.  We were able to obtain a good timing solution connecting back to our discovery observation, as shown in Figure~\ref{fig:resid} and given in Table~\ref{tab:timing}, \change{without rejecting any TOAs from the solution.} \change{We initially fit only for the rotation frequency $f$, but then added the frequency derivative $\dot f$ once  it was warranted by the residuals; the position was held fixed at the position from ATCA imaging.  We verified that this solution was robust and unique using \texttt{Algorithmic Pulsar Timing for Binaries} \citep[APTB;][]{2023arXiv231010800T}, which identified the same solution as we did by hand as the only solution.  This solution has a reduced $\chi^2$ of 1.97, suggesting that the uncertainties may be slightly underestimated or that there is some timing noise present.  Regardless, we leave the uncertainties in Table~\ref{tab:timing} as the unscaled values reported by \texttt{PINT} but note that they too may be slightly underestimated.  There is also a contribution to the uncertainties on $f$ and $\dot f$ from our uncertainties in the ATCA imaging position, which is not taken into account by \texttt{PINT}.  Using Appendix~A of \citet{2008ApJ...675.1468H}, we find  uncertainties of \Change{$6\times 10^{-10}\,$Hz} and $1.3\times 10^{-16}\,{\rm Hz\,s}^{-1}$ on $f$ and $\dot f$  from the position uncertainties, comparable to those from the timing analysis.  In the future, once the data span exceeds 1\,yr, we will be able to fit for a more precise timing position and reduce those uncertainties further.}

We find that the pulsar is young, with a characteristic age of only 34.6\,kyr.    Given the modest timespan of the current data, we cannot determine useful constraints on any higher timing derivatives.  Allowing a fit for the second frequency derivative $\ddot f$ finds $\ddot f=-(1.2\pm0.9)\times 10^{-22}\,{\rm Hz\,s}^{-2}$, consistent with 0, and implying a 3$\sigma$ upper limit on the braking index of $n\equiv f \ddot f/\dot f^2<102$. The distance to the pulsar is highly uncertain, with values from 4.3\,kpc \change{\citep[YMW16 model, ][]{2017ApJ...835...29Y}} to $>50\,$kpc \change{\citep[NE2001 model,][]{2002astro.ph..7156C}} depending on the Galactic electron density model.

\begin{deluxetable}{l l}
\tablecaption{Measured and derived parameters for \psr.\label{tab:timing}}
        \tablehead{
        \colhead{Parameter} & \colhead{Value}}
        \startdata
Right Ascension (J2000)\tablenotemark{a} & $10^{\rm h}32^{\rm m}50\fs54(1)$\\
Declination (J2000)\tablenotemark{a} & $-58{\degr}04\arcmin34.9(2)\arcsec$\\
Start (MJD) & 59999.6\\
End (MJD) & \change{60219.1}\\
Frequency (Hz) & \change{12.6998621581(4)}\\
Frequency Derivative (Hz\,s$^{-1}$) & $-5.81654(10)\times10^{-12}$\\
Epoch of Period (MJD) & \change{60100}\\
$\chi^2$/DOF & \change{45.5/23}\\
\change{RMS residual ($\mu$s)} & \change{504.7 }\\
Dispersion Measure (pc\,cm$^{-3}$) & 819$\pm$4\\
Rotation Measure (rad\,m$^{-2}$) & $-2000$ $\pm$ 1\\ \hline
Galactic Longitude (deg) & 285.436\\
Galactic Latitude (deg) & +0.008\\
Period (s) &  \Change{0.078741012111(2)}\\
Period Derivative (s\,s$^{-1}$) & \Change{$3.60634(6)\times10^{-14}$}\\
Characteristic Age (kyr) & 34.6\\
Surface Magnetic Field (G) & $1.7\times10^{12}$\\
Spin-down Luminosity (erg\,s$^{-1}$) & $2.9\times10^{36}$\\
Distance (kpc) & 4.3\tablenotemark{b}\\
 & $>50$\tablenotemark{c}\\
 \enddata
\tablecomments{Quantities in parentheses are 1$\sigma$ uncertainties on the last digit, \change{without any additional scaling}. \change{ The JPL DE405 solar system ephemeris has been used and times refer to TDB (using TT = TAI+32.184 s).}}
 \tablenotetext{a}{Derived from ATCA imaging.}
 \tablenotetext{b}{From the \citet{2017ApJ...835...29Y} electron density model.}
 \tablenotetext{c}{From the \citet{2002astro.ph..7156C} electron density model.}
\end{deluxetable}

\begin{figure}
    \includegraphics[width=0.98\columnwidth]{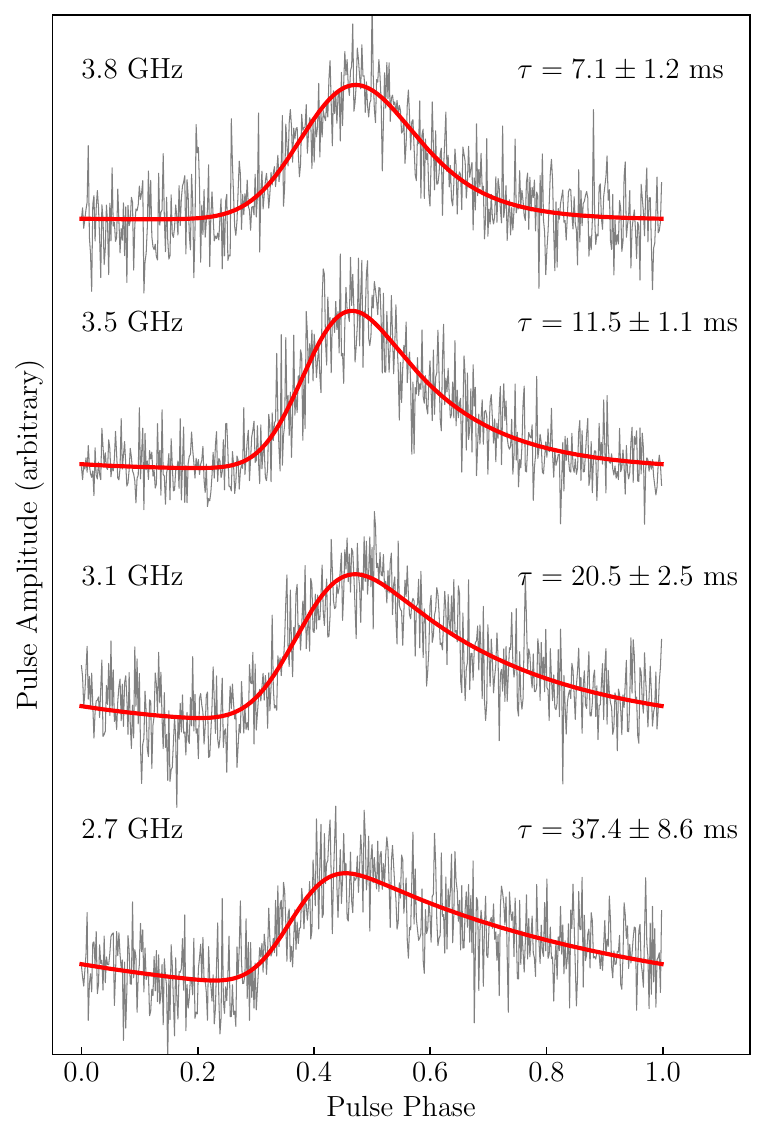}
    \caption{Normalized pulse profiles created from the summed Parkes UWL data of \psr\ as a function of observing frequency. We show profiles (gray) at 3.8, 3.5, 3.1 and 2.7\,GHz vertically offset from top to bottom, along with the best-fit exponential scattering model fits (red).  The timescales for the scattering fits are given. A power-law fit to the frequency dependence of $\tau$ gives $\alpha_s=-4.7\pm0.7$ (with $\tau(\nu)\propto \nu^{\alpha_s}$) and a scattering timescale at 3\,GHz of $22\pm2$\,ms. }
    \label{fig:scattering}
\end{figure}

\begin{figure}[hbt!]
    \includegraphics[width=0.80\columnwidth]{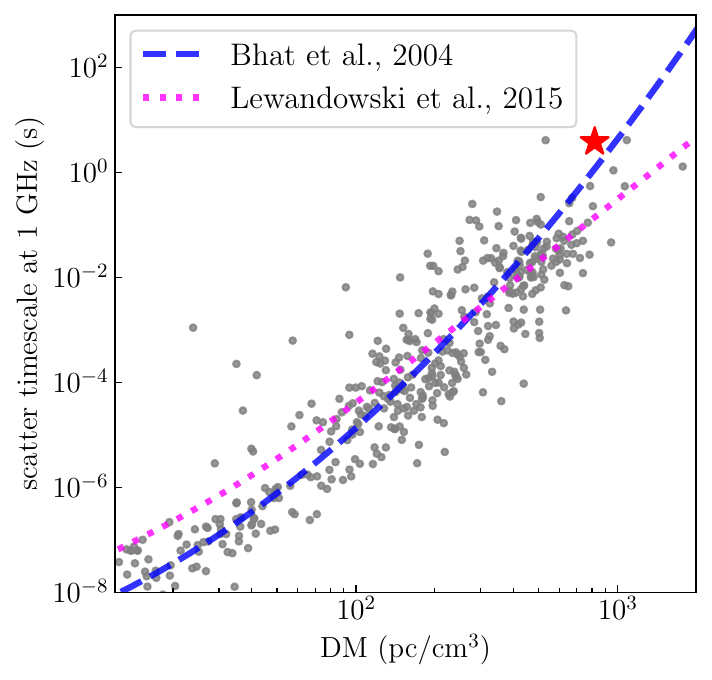}
    \caption{Scattering timescale as a function of the dispersion measure for \psr\ (red star) and pulsars (gray dots) in the ATNF pulsar catalog \citep[v.\ 1.70,][]{2005AJ....129.1993M}. The dashed line corresponds to the scattering timescale and DM relation fitted by \citet{2004ApJ...605..759B} while the dotted line corresponds to the one fitted by \citet{2015MNRAS.449.1570L}.}
    \label{fig:tauDM}
\end{figure}

Once we had a timing solution, we co-added all of the available Parkes/Murriyang data into a single high-S/N dataset. We then fit for an exponential scattering model \citep[e.g.,][]{2012hpa..book.....L}.  We assumed that the underlying profile model was a Gaussian, and that scattering convolved this with a frequency-dependent exponential.  Dividing up the data above $2.5\,$GHz into 4 sub-bands, we obtained the fits in Figure~\ref{fig:scattering}.  The scattering timescale for each sub-band was fit independently: when we fit for a power-law frequency dependence we obtain a spectral index $\alpha_s=-4.7\pm0.7$ (with $\tau(\nu)\propto \nu^{\alpha_s}$), consistent with expectations from Kolmogorov turbulence \citep{1986ApJ...310..737C,1986MNRAS.220...19R,1998ApJ...507..846C,2004ApJ...605..759B}, and a scattering timescale at 3\,GHz of $22\pm2$\,ms.  
The implied scattering timescale at 1\,GHz was 3845\,ms ($>40\times$ the pulse period),  a factor of 3 greater than that predicted by \citet{2004ApJ...605..759B} based on the DM, a factor of 27 greater than that predicted by \citet{2015MNRAS.449.1570L}, and a factor of 28 greater than that predicted by \citet{2017ApJ...835...29Y}\footnote{The model of \citet{2002astro.ph..7156C} is not sensible along this line-of-sight for DM $>$ 620\,pc\,cm$^{-3}$.}. 
However, the scattering timescale of \psr\ is still largely consistent with the whole pulsar population (see Figure~\ref{fig:tauDM}) considering the large internal uncertainties of these models. 

\change{Using the full Stokes fold mode (coherently dedispered) UWL observations we also conducted flux and polarization calibration. For polarization calibration, we used a short $\sim$2-min observation of a linearly polarized noise diode, which was obtained at the start of each observing session. We used the radio Galaxy PKS B0407--658 as a flux density reference, which is a more reliable calibration source for the UWL. Prior to performing any analysis all observations were manually cleaned from RFI. For both polarimetric and flux calibration of pulsar observations, we used \texttt{PSRCHIVE's} \texttt{pac} pulsar archive calibration program which was first used to generate a calibrator database. This database file was then used to obtain a flux and polarization calibrated file. This methodology of flux and polarization calibration is similar to the one outlined in \citet{lower_20} and references therein.}

\begin{figure}
    \includegraphics[width=0.98\columnwidth]{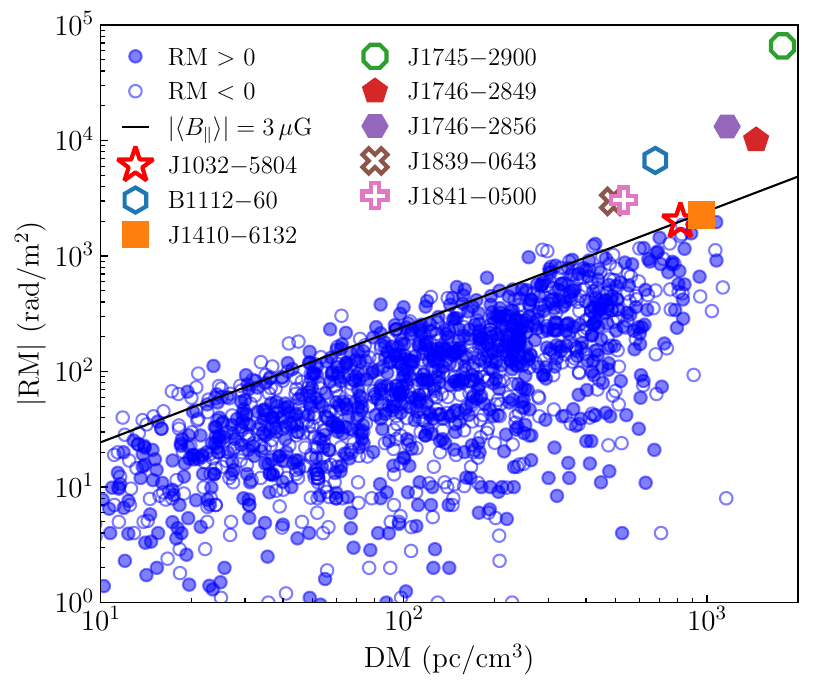}
    \caption{Dispersion measure (DM) vs.\ rotation measure (RM) for sources in the ATNF pulsar catalog \citep[v.\ 1.70,][]{2005AJ....129.1993M}.  Sources with positive RM are filled, and those with negative RM are empty.  Individual sources with $|{\rm RM}|>2000\,{\rm rad\,m}^{-2}$ are labeled, and \psr\ is the red star.  We also plot a line indicating $|{\rm RM}|/{\rm DM}=2.44\,{\rm rad}\,{\rm cm}^3\,{\rm pc}^{-1}\,{\rm m}^{-2}$, corresponding to an average interstellar magnetic field along the line-of-sight of $3\,\mu$G.}
    \label{fig:RM}
\end{figure}

\begin{figure}[hbt!]
    \centering
    \includegraphics[width=0.95\columnwidth]{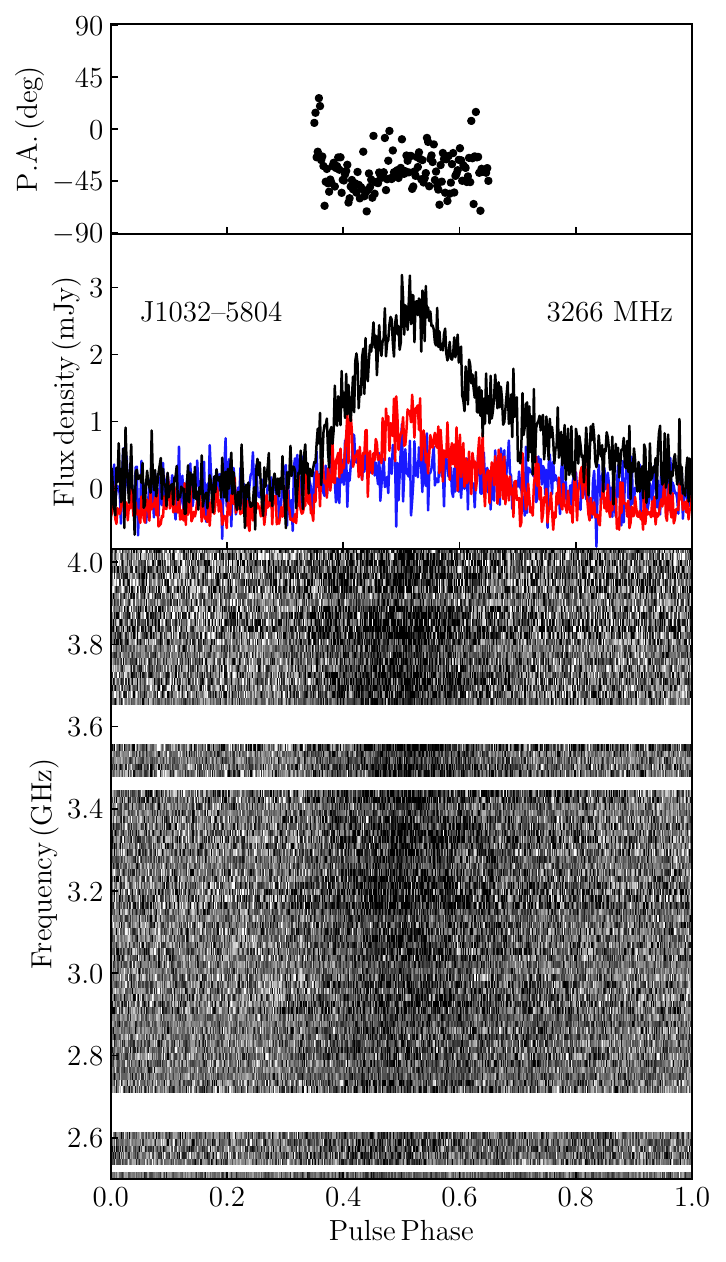}
    \caption{Dynamic spectra and polarization profiles of PSR J1032$-$5804 from a 52\,min observation using the Parkes UWL receiver on 2023~May~02. The bottom panel shows the dynamic spectrum over the frequency range of 2.5--4\,GHz  in which pulsar is visible. The horizontal stripes correspond to the parts of the band that have been removed due to the presence of RFI. The middle panel shows the frequency-averaged profiles of the total intensity (in black), linear polarization (in red), and circular polarization (in blue). The top panel shows the polarization position angle (P.A) at a central frequency of 3.25\,GHz.}
    \label{fig:polarimetry}
\end{figure}

We then used the \texttt{rmfit} tool of \texttt{PSRCHIVE} to fit for the rotation measure (RM) in one of the highest signal-to-noise ratio observations. We find a strong detection\footnote{\Change{We used an RM search range between $-3000$ to 3000 rad\,m$^{-2}$ to obtain the best \texttt{rmfit} value.}} with ${\rm RM}=-2000 \pm 1\,{\rm rad\,m}^{-2}$. 
Though the RM measurement of \psr\ is almost three times the most extreme $|{\rm RM}|$ of known pulsars within 5 degrees of \psr, the average interstellar magnetic field along the line-of-sight $B=1.232\,{\rm RM}/{\rm DM}\approx-3\,\mu$G is consistent with the Galactic magnetic field model fitted by \citet{2018ApJS..234...11H}, and also with the Galactic pulsar distribution (see Figure~\ref{fig:RM}).
Based on the pulse profiles (see Figure~\ref{fig:polarimetry}), we find a circular polarization fraction of 10\% at a frequency of 3\,GHz, which is not consistent with the value of $\sim$25\% found from ASKAP imaging at 843\,MHz. 
\Change{We also check the polarization evolution at three UWL frequencies (2.7, 3.2, 3
8\,GHz) and find that the circular polarization fraction remains similar across these three UWL bands.}
However, we caution that the very broad pulse profile makes establishing a reliable pulse baseline -- and therefore the total-intensity pulsed flux density -- difficult.  We also find that there is very little position-angle change across the pulse, likely due to the long scattering timescale even at these frequencies \change{\citep[e.g.,][]{2003A&A...410..253L,2015A&A...576A..62N}}.

Apart from examining the highest S/N observation, we also examined other observations taken at different epochs and found discrepancies in the RM values of up to 50\,rad\,m$^{-2}$ between them. These variations might resemble the changes in the magneto-ionic conditions observed in pulsars within the Galactic Center \change{\citep[albeit on a smaller scale;][]{2018ApJ...852L..12D,2023MNRAS.524.2966A} or elsewhere \citep{2021MNRAS.502.1253J}}. They could also be due to instrumental effects. \change{Furthermore, different levels of RFI excision and changing SNRs coupled with the highly scattered pulse profiles can potentially cause these discrepancies.} Our ongoing investigation aims to ascertain whether these fluctuations stem from instrumental factors, determine the time frame of these variations, and establish any potential correlations with changes in DM. This comprehensive analysis is the subject of future work.



\subsection{Radio Continuum}\label{subsec:radio}

\begin{figure}
    \centering
    \includegraphics[width=0.95\columnwidth]{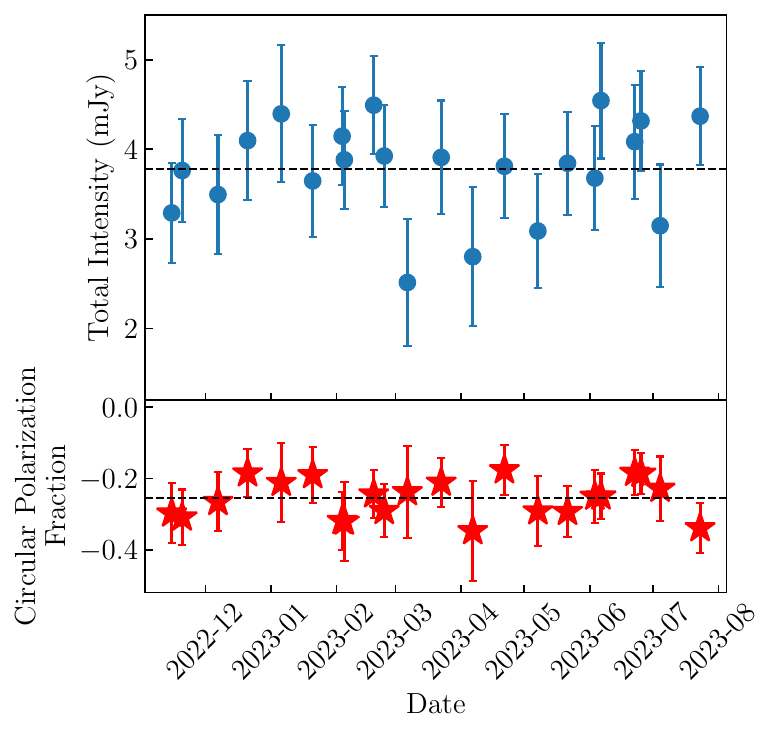}
    \caption{\change{Total intensity (Stokes I) flux densities of \psr\ from the VAST survey (top), along with circular polarization fractions (bottom).  Both are consistent with constant values (dashed lines).  We find a mean Stokes I flux density of $3.82\pm0.13$\,mJy, and a mean circular polarization fraction of $-0.24\pm0.02$. }}
    \label{fig:vastlc}
\end{figure}

We looked at all available epochs of ASKAP data from VAST (22 observations so far) to probe the flux variability of \psr\ \change{(see Figure~\ref{fig:vastlc})}.  We used forced flux measurements\footnote{\change{We performed the fitting at the source position with the synthesized beam of that epoch.}} to deal with non-detections.  
Both the flux density and circular polarization fraction are consistent with no variability, with $\chi^2$ of \change{$14.7$}  for the flux density and \change{$10.8$} for the polarization fraction, both for \change{21} degrees-of-freedom.

We also checked other ASKAP surveys from the CSIRO ASKAP Science Data Archive (CASDA).
\psr\ was detected in the RACS-mid survey \citep{2023PASA...40...34D} with a peak flux density of $2.84\pm0.38\,$mJy\,beam$^{-1}$ at 1367.5\,MHz.
It was also detected in the Evolutionary Map of the Universe (EMU; \citealt{2011PASA...28..215N, 2021PASA...38...46N}) project with a peak flux density of $5.16\pm0.36\,$mJy\,beam$^{-1}$ and $4.96\pm0.37\,$mJy\,beam$^{-1}$ in the EMU fields 1017$-$60 (Schedule Block 46953) and 1029$-$55 (Schedule Block 46915) at 943.5\,MHz, respectively\footnote{The EMU observations are typically $\sim$10\,hrs, and the high flux uncertainties are mainly from the extended emission nearby.}.

We also checked the GaLactic and Extragalactic All-sky Murchison Widefield Array \citep[GLEAM;][]{2015PASA...32...25W, 2017MNRAS.464.1146H} survey data.
There were no detections of the pulsar in any GLEAM bands, with a 3$\sigma$ upper limit of 1.62\,Jy\,beam$^{-1}$, 0.78\,Jy\,beam$^{-1}$, 0.28\,Jy\,beam$^{-1}$, and 0.14\,Jy\,beam$^{-1}$ at 88\,MHz, 118\,MHz, 154\,MHz, and 200\,MHz, respectively. The elliptical structure in the ASKAP images is visible, albeit at low resolution and significance.

Unlike most pulsars whose radio spectra can be modeled by a simple power law \citep[e.g.,][]{2013MNRAS.431.1352B}, the spectrum of \psr\ may  peak at $\sim$1\,GHz and decline at higher and lower frequencies.
To obtain the spectral energy distribution over a wide frequency range, we observed \psr\ with the Australian Telescope Compact Array (ATCA) on 2023 Aug 21 (project code C3363) with 6D configuration in L (1--3\,GHz) and C/X (5--7\,GHz and 8--10\,GHz)  bands for three hours each.
The observations were calibrated using \object[PKS B1934-638]{PKS~B1934--638} for the flux density scale and the instrumental bandpass. \object[PMN J1047-6217]{PMN~J1047--6217} was used for phase calibration.
We used \textsc{Casa} to perform both the data calibration and the continuum imaging.
We split each band into two parts to image the data.
We detected a compact radio source at the pulsar position in all six subbands (see Table~\ref{tab:atca_radio}).
The best-fit position of \psr\ from the ATCA observation (based on the X-band data) is RA $10^{\rm h}32^{\rm m}50\fs54\pm0\fs01$, Dec $-58{\degr}04\arcmin34\farcs9\pm0\farcs2$, which is within the ASKAP $1\sigma$ positional error circle.


\begin{deluxetable}{c c c}
\tablecaption{ATCA flux density measurement for \psr. Non-detections are denoted by 3$\sigma$ upper limits based on the local noise. }
\label{tab:atca_radio}
\tablecolumns{3}
\tablehead{
\colhead{Frequency Range} &  
\colhead{$S_{\rm Stokes I}$} &
\colhead{$S_{\rm Stokes V}$} \\ 
\colhead{(MHz)} & 
\colhead{(mJy/beam)} &
\colhead{(mJy/beam)}
} 
\startdata
1076--2100 & $2.09\pm0.64$ & $-0.56\pm0.07$ \\
2100--3124 & $1.76\pm0.20$ & $-0.19\pm0.06$ \\
4476--5500 & $0.73\pm0.04$ & $<0.14$ \\
5500--6524 & $0.56\pm0.05$ & $<0.13$ \\
7976--9000 & $0.22\pm0.03$ & $<0.10$ \\
9000--10024 & $0.19\pm0.03$ & $<0.12$
\enddata
\end{deluxetable}

\subsection{Swift X-ray Observations}
Motivated by the fact that \psr\ seems to be young and energetic, we looked for available X-ray data but there were no observations with \textit{Chandra} or \textit{XMM-Newton} that covered this location.  There was one short observation with the \textit{Neil Gehrels Swift Observatory} \citep{2004ApJ...611.1005G}, but the duration was only 400\,s.  Therefore we requested a longer  Director's Discretionary Time observation.  We observed \psr\ using the \textit{Swift}    X-ray Telescope (XRT; \citealt{2005SSRv..120..165B}) for a total of 2242\,s in two exposures on 2023~May~22 and 2023~May~26.  We retrieved the merged data-set created with the online analysis tools\footnote{\url{https://www.swift.ac.uk/user_objects/}.}.  We find 0 events within $15\arcsec$ of \psr, and derive a 95\% upper limit on the count-rate of $5\times 10^{-4}\,{\rm s}^{-1}$.

To interpret the X-ray upper limit, we first convert the DM into a hydrogen column density using the relationship of \citet{2013ApJ...768...64H}, and find $N_{\rm H}\approx 2.5\times 10^{22}\,{\rm cm}^{-2}$.  We assume a power-law X-ray spectrum with index $\Gamma=1.5$, typical of young pulsars like \psr\ \citep[][although \citealt{2003ApJ...591..361G} would predict a flatter spectrum for this $\dot E$]{2007ApJ...665.1297H,2008AIPC..983..171K, 2012ApJS..201...37K}.  This then gives an upper limit to the unabsorbed flux (0.5--8\,keV) of $5.5\times 10^{-14}\,{\rm erg\,cm}^{-2}\,{\rm s}^{-1}$, or an upper limit to the luminosity of $1.6\times 10^{32}d_5^2\,{\rm erg\,s}^{-1}$, where the distance is $5d_5\,$kpc.  We can compare this with the spin-down luminosity and find $L_X/\dot E < 5\times 10^{-5}d_5^2$.  This is lower than many young pulsars but not inconsistent with the tail of the population \citep{2008AIPC..983..171K, 2012ApJS..201...37K}.  A deeper X-ray observation may be able to more robustly constrain any X-ray emission, although the highly-uncertain distance will make any constraints somewhat weak.


\section{Discussion}\label{sec:discuss}


\psr\ was detected as a point source in all ASKAP and ATCA observations. Despite the large scattering timescale, the angular broadening effect is small compared to the synthesized beam of all images. For a source at a distance $d$ scattered by a single thin screen at a distance $s$, the expected full width at half maximum of angular broadening follows
$$
\theta = \sqrt{\frac{8\ln2c\left(d-s\right)\tau}{ds}}
$$
where $\tau$ is the scattering timescale and $c$ is the speed of light \citep{1997ApJ...475..557C}.
Assuming $d=4.3\,$kpc, and $s=d/2$, the expected angular broadening is $\theta\approx2\farcs1\left(\nu/1\,{\rm GHz}\right)^{-2.2}$.
All synthesized beam sizes are much larger than the expected value at corresponding frequencies (e.g., $\theta_{2.1\,{\rm GHz}} \approx 0\farcs4$, while the ATCA synthesized beam at 2.1\,GHz is $\sim5\arcsec$).
The large scattering timescale also leads to low-level time variability in the image domain.
Assuming a Kolmogorov spectrum \citep[e.g.,][]{1977ARA&A..15..479R}, we calculated a diffractive scintillation bandwidth of $\Delta f_{\rm DISS} \ll 1\,$kHz and a scintillation timescale of $\ll 10\,$s at 888\,MHz.
Both the scintillation bandwidth and the timescale are much smaller than those for typical VAST observations (288\,MHz and 12\,mins), which could explain the non-variability of the source in the VAST survey.

\begin{figure*}[hbt!]
    \centering
    \includegraphics[width=\textwidth]{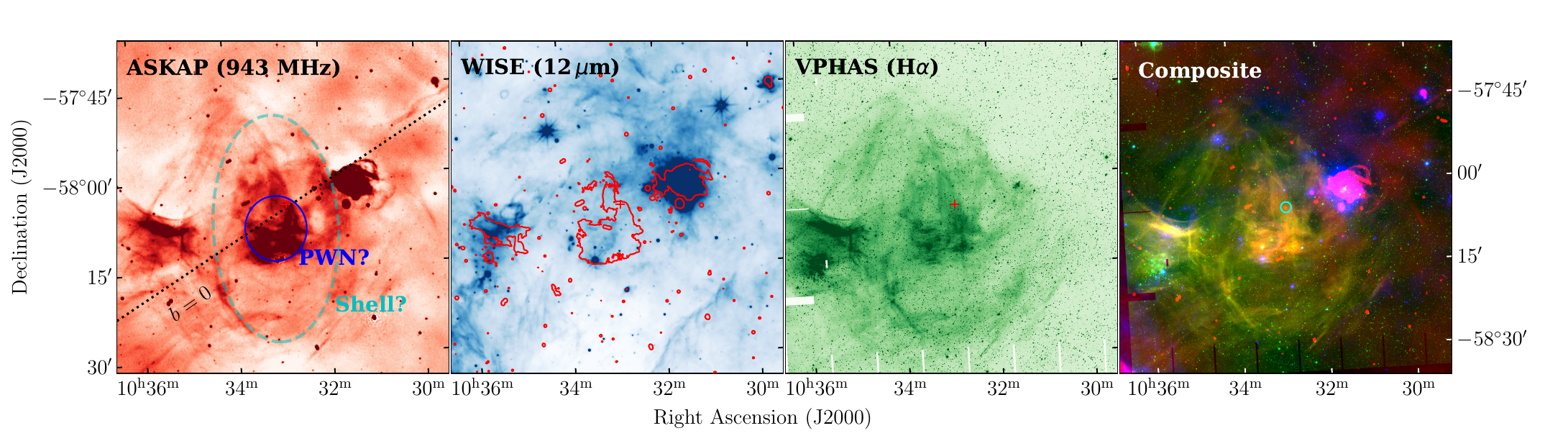}
    \caption{Field of \psr, observed with ASKAP at 943\,MHz (from EMU, \citealt{2011PASA...28..215N}; left), \textit{WISE} (from the ALLWISE data release; \citealt{2013wise.rept....1C}) 12\,$\mu$m (middle left), and H$\alpha$ from the VST Photometric H$\alpha$ Survey (VPHAS; \citealt{2014MNRAS.440.2036D}) (middle right).  All images are roughly $0.8\degr$ on a side with north up.  The position of \psr\ is shown by the red plus.  The Galactic plane is indicated with the black dotted line.  On the ASKAP image, we show the outline of a potential SNR shell (cyan dashed ellipse) and a more compact pulsar wind nebulae (magenta circle).  On the \textit{WISE} image we also show contours from the ASKAP image (red).
    We also showed a composite of radio (EMU, red), infrared (\textit{WISE}, blue), and H$\alpha$ (VPHAS, green) emission (right), where the position of \psr\ is shown by the cyan circle.
    }
    \label{fig:images}
\end{figure*}

Motivated by the youth of \psr, we searched for an associated supernova remnant (SNR). There are no cataloged SNRs coincident with the pulsar, with the  closest  $>1.5\degr$ away \citep[][2022 December version]{2019JApA...40...36G}.  We therefore examined deeper radio images from the EMU project\footnote{The ATCA images from Section~\ref{subsec:radio} were not useful in searching for extended emission because of limited $uv$ coverage.}. 
The 10-hour EMU observation reaches a sensitivity of ~28 $\mu$Jy\,beam$^{-1}$ with a resolution of 15$^{\prime\prime}$.
We retrieved EMU field 1017$-$60 from CASDA.  Figure~\ref{fig:images} shows a cutout of the region around \psr.  There is a considerable amount of extended emission around the position of \psr\ in the EMU image, with a potential SNR shell outlined in cyan and a more compact emission region (potentially a pulsar wind nebula or PWN) outlined in blue.  However, we caution that there is a lot of other extended emission in this region and it is not confined to only the area around \psr, suggesting that there could be contributions from Galactic \ion{H}{2} regions or unrelated synchrotron emission.  Unfortunately, difficulties in robustly deconvolving the complex extended emission in this region prevent us from establishing a reliable in-band spectral index that might help discriminate between these sources of emission.  Nor can we identify a radio image at another wavelength to do the same.  Instead we examined the 12\,$\mu$m image from ALLWISE \citep{2013wise.rept....1C}.  We also assembled an H$\alpha$ mosaic from 6 individual exposures from the VST Photometric H$\alpha$ Survey (VPHAS; \citealt{2014MNRAS.440.2036D}), which we mosaiced together with \texttt{swarp} \citep{2002ASPC..281..228B} to remove the gaps between the individual detectors.
These images are shown in Figure~\ref{fig:images}.  Plotting the EMU contours on the ALLWISE image, we see that most of the radio emission is roughly traced by the 12\,$\mu$m emission as expected for thermal bremsstrahlung emission from \ion{H}{2} regions \citep[e.g.,][]{1999ApJS..123..219C, 2016era..book.....C, 2022A&A...664A.140K}.  However, the central emission closest to \psr, the putative PWN, does not show any emission at 12\,$\mu$m.  That region is bright in H$\alpha$, and there is also a good correspondence between the diffuse radio emission (putative PWN) surrounding \psr\ and the diffuse H$\alpha$ emission, including narrow filaments (putative shell).  Other regions that are bright at 12\,$\mu$m are seen at radio and optical wavelengths, suggesting thermal emission, but the region surrounding the pulsar generally lacks mid-infrared emission. This suggests that it may in fact be non-thermal emission associated with \psr, but we will need better multi-wavelength imaging (including potentially deeper X-ray images) or optical spectroscopy to confirm this.

\begin{figure}[hbt!]
    \centering
    \includegraphics[width=0.95\columnwidth]{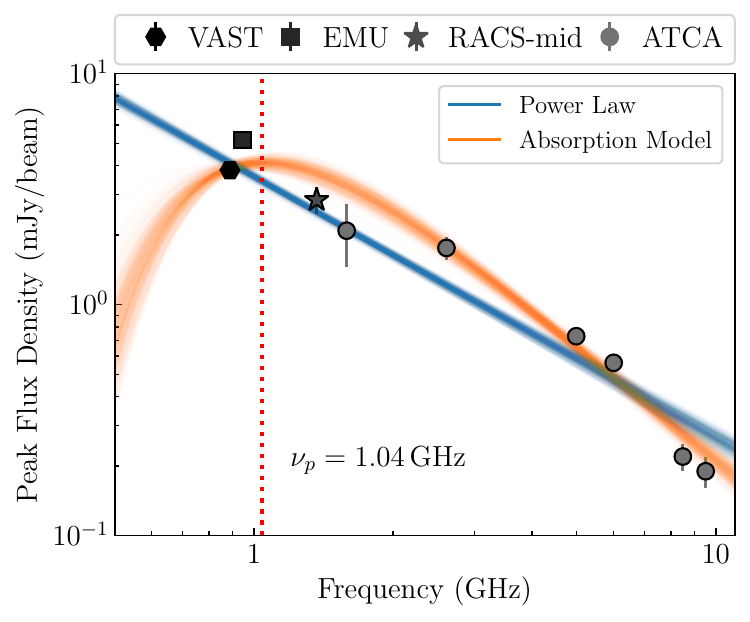}
    \caption{The radio spectrum of \psr. \change{Besides the ATCA data listed in Table~\ref{tab:atca_radio}, we also used VAST, EMU, and RACS-mid data in this fit. For the surveys with multiple observations (VAST and EMU), we used the mean flux density mentioned in Section~\ref{subsec:radio}.}
    The orange line shows the fitted free-free absorption model and the red dashed line shows the corresponding peak frequency, while the blue line shows the simple power law fit with a spectra index of $-1.13\pm0.03$.}
    \label{fig:sed}
\end{figure}

The SED of \psr\ is shown in Figure~\ref{fig:sed} (we excluded GLEAM data points here as they are not constraining).  The SED is ambiguous and noisy: it may continue as a typical pulsar power-law, but it also may peak at a frequency $\sim 1\,$GHz.  This may be  evidence of gigahertz-peaked spectrum (GPS) behavior.
Previous studies have suggested that the origin of the gigahertz spectral turnover is likely caused by the absorption from the pulsar ambient environment, such as supernova remnants (SNRs), pulsar wind nebula (PWNe), and \ion{H}{2} regions \citep[e.g.,][]{2011A&A...531A..16K, 2016MNRAS.455..493R}.
In Figure~\ref{fig:sed}, we fit the SED of \psr\ with a simple power-law and the thermal free-free absorption model described by \citet{2015ApJ...808...18L}
$$
S\left(\nu\right) = S_1\left(\nu/{1\,{\rm GHz}}\right)^\alpha e^{-B\nu^{-2.1}}
$$
where $S_1$ is the pulsar's intrinsic flux density at 1\,GHz, $\alpha$ is the pulsar's intrinsic spectral index, and $B = 0.08235\times T_e^{-1.35}{\rm EM}$ (emission measure ${\rm EM} = n_e^2 s$; $n_e$, $s$, and $T_e$ are the absorbing material electron density, size, and electron temperature, respectively). For the power-law model, we found a spectral index of $-1.13\pm0.03$.  For the absorption model we estimated $S_1 = 9.61^{+1.52}_{-1.19}$, $\alpha = -1.66^{+0.08}_{-0.08}$, and $B=0.86^{+0.12}_{-0.12}$, and found the peak frequency $\nu_p = 1.04\,$GHz.
The peak frequency for \psr\ is consistent with the known GPS pulsar population (see Figure~7 in \citealt{2021ApJ...923..211K}).
We also note that there are deviations from both the free-free absorption model and the power-law model. We compared these two models using the $F$-statistic \citep[e.g.,][]{weisberg2005applied}. 
The distribution of the $F$-statistic follows an $F$-distribution with $d_1=1$, $d_2 = 6$ degrees of freedom.
We calculated an $F$-statistic of 6.80 with a corresponding $p$-value of 0.04, which means that the free-free absorption model fits significantly better than the power law model.

\begin{figure}[hbt!]
    \centering
    \includegraphics[width=0.95\columnwidth]{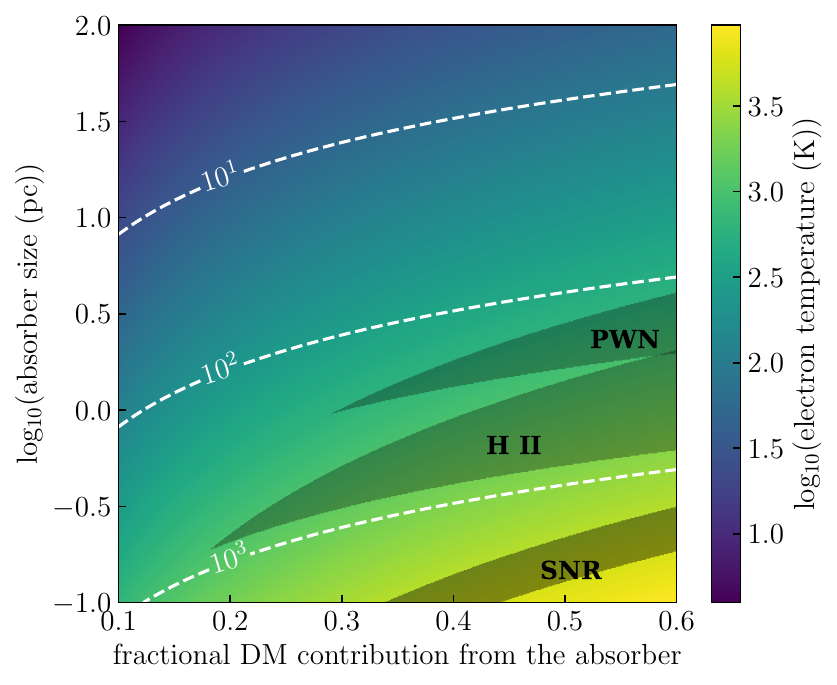}
    \caption{Constraints on the electron temperature $T_e$ (white dashed lines) and electron density $n_e$ (heat map) based on the best-fitted parameters (i.e., $B=0.99$). The x-axis shows the fractional DM contribution from the absorber, and the y-axis shows the size of the absorber on a logarithmic scale. 
    We show expected values for different absorbers in shaded regions: 1) SNR filament with $n_e \gtrsim 10^3\,$cm$^{-3}$ and $T_e \sim 5000\,$K \citep[e.g.,][]{2013ApJ...770..143L}; 
    2) PWN with $n_e \sim 50-250\,$cm$^{-3}$ and $T_e\sim1500$\,K \citep[e.g.,][]{2006ARA&A..44...17G}; 
    3) \ion{H}{2} region with $n_e \gtrsim 10^2\,$cm$^{-3}$ and $T_e\sim1000-5000\,$K. \citep[e.g.,][]{2006MNRAS.372..457S}.
    The typical sizes of an SNR filament, PWN, and \ion{H}{2} are $0.1\,$pc, $1\,$pc, and $10\,$pc, respectively. Assuming a free-free absorption model, our result largely agrees with the SNR filament or PWN scenario. 
    }
    \label{fig:absorb_prop}
\end{figure}

In Figure~\ref{fig:absorb_prop}, we show the constraints on the electron density and temperature of the absorber based on the fitted parameters and considered three scenarios: a dense SNR filament (with $s=0.1\,$pc), a PWN (with $s=1.0\,$pc), and a cold \ion{H}{2} region (with $s=10.0\,$pc).
The results are broadly consistent with either the SNR filament or the PWN scenario (see \citealt{2021ApJ...923..211K}, and references therein), which agrees with the EMU extended emission analysis above.
For both scenarios, the expected fractional DM contribution from the local environment is around 50\%, which could potentially break the inferences we made in previous discussions regarding the correlations between DM, RM, and scattering timescale.
For example, with a large DM contribution, the magnetic field of the local environment could dominate the RM value we measured, which could lead to a wrong estimation of the average interstellar magnetic field along the line of sight.
The relation between scattering timescales and DMs could also be significantly affected if there are additional screens near the pulsar \citep[e.g.,][]{2016ApJ...817...16C}.

Besides free-free absorption, synchrotron self-absorption in the pulsar magnetosphere \citep{1973A&A....28..237S} or  flux dilution caused by anomalous scattering \citep{2001ApJ...549..997C, 2014MNRAS.445.3105D} can also lead to a decrease in the flux density at lower frequencies.
Given that the data we have are all above $\sim$0.8\,GHz, it is hard to get a robust constraint on the surrounding environment for \psr\ without low-frequency observations.
Further deep radio continuum observations (especially at lower frequencies) of the pulsar itself and the surrounding extended emission and/or deeper X-ray images may be helpful in understanding the nature of the pulsar spectrum and probing the interstellar medium in the pulsar's local environment (e.g., potential SNR association).

\begin{figure}[hbt!]
    \includegraphics[width=0.98\columnwidth]{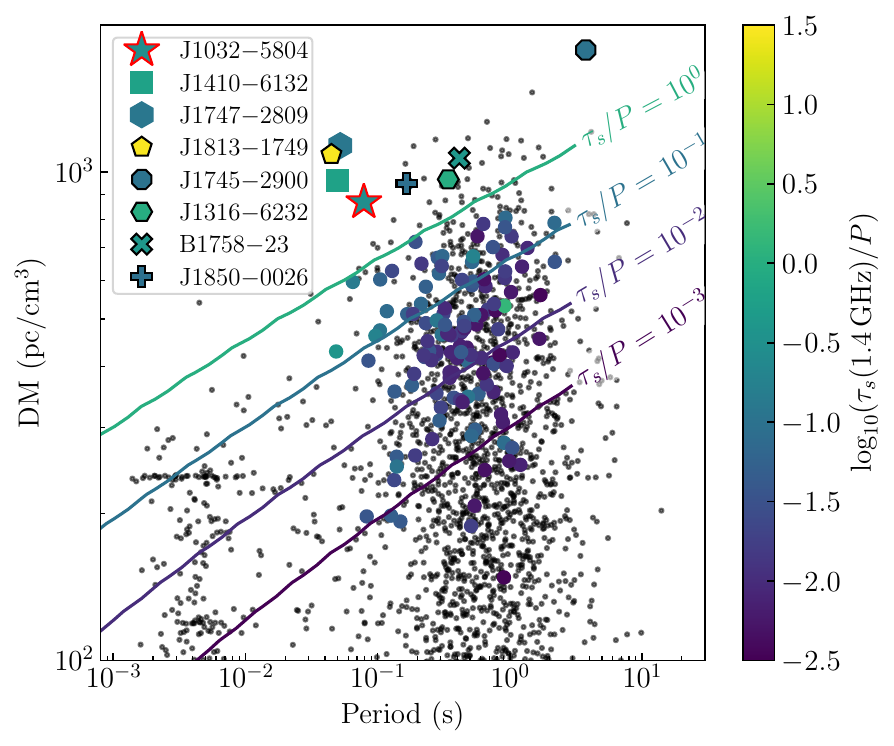}
    \caption{Dispersion measure vs.\ pulse period $P$ for sources in the ATNF pulsar catalog (v. 1.70).  Pulsars without cataloged scattering timescales are shown as black points.  Those with cataloged scattering timescales $\tau_s$ at 1.4\,GHz are colored according to the ratio of the timescale divided by the pulse period, with the color-bar shown on the right.  We also plot contours of constant $\tau_s/P$ based on the model of \citet{2004ApJ...605..759B}.  Sources with $\tau_s/P>1$ should not be detectable as pulsed sources at 1.4\,GHz.  Individual highly-scattered pulsars are labeled, with values for PSR~J1410$-$6132 taken from \citet{2008MNRAS.388L...1O} and PSR~J1747$-$2809 taken from \citet{2009ApJ...700L..34C}.  \psr\ is the star with the red outline.}
    \label{fig:P_DM}
\end{figure}

\psr\ is the third most scattered pulsar known, which makes it hard to detect in periodicity searches. 
In Figure~\ref{fig:P_DM} we plot the DM vs.\ period for the pulsars in the ATNF pulsar catalog \citep{2005AJ....129.1993M}.  In general, as DM increases so does the scattering \citep[][]{2002astro.ph..7156C,2004ApJ...605..759B,2015MNRAS.449.1570L, 2017ApJ...835...29Y}.  Those sources with the highest DMs and the shortest periods will then be the hardest to detect in a pulsation search.  This is illustrated not only by the case of \psr, but by other similar sources such as PSR~J1813$-$1749 \citep{2021ApJ...917...67C,2009ApJ...700L.158G,2012ApJ...753L..14H,2010RMxAA..46..153D}.  While PSR~J1813$-$1749 is even more scattered than \psr, both of them are undetectable as pulsed sources at the normal search frequency of 1.4\,GHz.  This is consistent with the fact that the location of \psr\ was searched as part of the Parkes multi-beam pulsar survey \citep{2001MNRAS.328...17M} and the deeper High Time Resolution Universe mid-latitude survey \citep{2010MNRAS.409..619K,2011MNRAS.416.2455B}, both centered at 1.4\,GHz, but it was not identified.

Many previous pulsar surveys are not sensitive to the highly scattered short-period pulsars as most of them were conducted at lower frequencies (ranging from $\sim$300 to 1400\,MHz).
Searching for pulsars at high frequency is one of the ways to detect more new pulsars of this kind.
However, only a few new pulsars have been discovered in previous very high frequency ($\nu\gtrsim5\,$GHz) pulsar surveys \citep[e.g.,][]{2011MNRAS.411.1575B, 2021MNRAS.507.5053E, 2021A&A...650A..95T, 2022ApJ...933..121S}, which is possibly due to the limited sensitivity (flux densities of pulsars will be low at high frequencies due to their steep radio spectra) and insufficient sky coverage (small beam size at high frequency makes it hard to tile the sky efficiently).
With new high-frequency pulsar surveys, such as MMGPS at S-band (1.7--3.5\,GHz), more scattered pulsars are expected to be discovered \citep{2023MNRAS.524.1291P}.
To roughly quantify this, we used \texttt{PsrPopPy} \citep{2014MNRAS.439.2893B} \footnote{\url{https://github.com/samb8s/PsrPopPy}} to model the millisecond pulsar (MSP) and normal pulsar populations.
We found that $\sim$20\% ($\sim$5\%) of MSPs (normal pulsars) could be missed in pulsar surveys at 1.4\,GHz (assuming the sensitivity to be $\sim$0.1\,mJy) due to scattering (i.e., scattering timescale longer than the period).
When we change the central frequency to $3\,$GHz, $\sim$70\% of pulsars will be undetectable due to the lower flux density (assuming that the sensitivities for 1.4\,GHz and 3\,GHz surveys are both 0.1\,mJy), but $\sim$70\% (90\%) of previously undiscovered MSPs (normal pulsars) will be visible to the 3\,GHz survey, though the computational cost could be prohibitively expensive due to the need to \change{search for a large number of pointings}.

The prospects for new pulsar discovery in imaging domain are promising with the development of new radio continuum surveys, including the GaLactic and Extragalactic All-Sky MWA Extended Survey (\citep[GLEAM-X][]{2022PASA...39...35H}, the LOFAR Two-metre Sky Survey \citep[LoTSS;][]{2017A&A...598A.104S}, the ASKAP Rapid Continuum Survey (RACS; \citealt{2020PASA...37...48M, 2023PASA...40...34D}), the Evolutionary Map of the Universe survey with ASKAP\citep[EMU;][]{2011PASA...28..215N, 2021PASA...38...46N}, the Karl G. Jansky Very Large Array (VLA) Sky Survey \citep[VLASS;][]{2020PASP..132c5001L}, and The HUNt for Dynamic and Explosive Radio transients with MeerKAT \citep[THUNDERKAT;][]{2016mks..confE..13F} survey.
For ASKAP surveys, we considered shallow surveys ( with $\sim$12\,mins integration time achieving a typical detection threshold of $\sim$1.5\,mJy) and deep surveys (with $\sim$10\,hrs integration time achieving a typical detection threshold of $\sim$50\,$\mu$Jy), and quantified the potential new detections with \texttt{PsrPopPy} \citep{2014MNRAS.439.2893B}.
We expected to detect 4 (8) highly scattered \footnote{Here we only considered the pulsars that are undetectable with a 1.4\,GHz pulsar survey. Some of the pulsars discussed below are expected to be detected in a 3\,GHz pulsar survey as well.} MSPs (normal pulsars) in shallow surveys, and 483 (600) in deep surveys.
Assuming $\sim$10\% of pulsars are detected with circularly polarized emission \citep[e.g.,][]{2018MNRAS.474.4629J}, the number of highly scattered MSPs (normal pulsars) that can be detected via circular polarization searches alone with ASKAP is expected to be 1 (1) for shallow surveys and 48 (60) for deep surveys.
It is unlikely for shallow surveys to discover a large number of highly scattered pulsars with circular polarization searches, but using deep surveys or stacking several shallow surveys can potentially discover a handful of new scattered pulsars.
Besides circular polarization searches, other image domain techniques, such as searching for steep spectrum sources, high energy sources (e.g., X-ray, $\gamma$-ray) associations, and potential SNRs (candidate) associations, can also be used in discovering extremely scattered pulsars.

\section{Conclusions}\label{sec:concls}

We discovered a young, highly scattered pulsar \psr\ in the Galactic plane in a search for circularly polarized sources as part of the ongoing ASKAP-VAST survey.
The pulsar has a period of 78.7\,ms and a DM of \change{$819\pm4\,$pc\,cm$^{-3}$}.
The long scattering timescale $\tau_{\rm 1\,GHz}\approx3.84\,$s makes it the third most scattered pulsar known and also explains the non-detection in previous pulsar surveys despite its high flux density.
Besides circular polarization, linear polarization emission was also detected in the follow-up observations with Murriyang/Parkes. The pulsar has an RM of \change{$-2000\pm1\,$rad\,m$^{-2}$}. Though the measured RM is among the highest RM detected, it is consistent with the general RM--DM trend, and also the Galactic large-scale magnetic field model \citep[e.g.,][]{2018ApJS..234...11H}.

\psr\ is young, with a characteristic age of $34.6$\,kyr. No X-ray emission was detected in \textit{Swift} observations, which gives an upper limit to the 0.5-8\,keV X-ray luminosity of $1.6\times10^{32}d^2_5\,$erg\,s$^{-1}$. The ratio of the X-ray luminosity to the spin-down luminosity is lower than many young pulsars, but is still consistent with the tail of the population. Further deeper X-ray observation may be able to constrain any X-ray emission from the pulsar itself, the PWN, or the SNR shell.
ATCA observations combined with the archival ASKAP observations revealed that \psr\ is a potential GPS source, which suggested strong absorption along the line-of-sight.
A preliminary analysis in this work based on the surrounding extended radio emission and the pulsar radio spectrum may infer the existence of the PWN or SNR.
Further observations to measure the spectral energy distribution of the pulsar itself and the spectral map of the extended emission are useful to understand its local environment, and hence probe the properties of the interstellar medium in the vicinity of the pulsar.

This discovery highlights the possibility of discovering new pulsars (especially extreme ones) from continuum images. We can identify more highly scattered pulsars like \psr\, with the high sensitivity and good resolution data from the ongoing ASKAP surveys.
In the future, with the construction of next-generation radio telescopes such as the Square Kilometre Array, the Deep Synoptic Array, and the Next Generation Very Large Array,  imaging domain searches will become a more  powerful tool for discovering extreme pulsars (e.g., highly accelerated, highly scattered, and highly intermittent) that are hard to find via traditional surveys.

\begin{acknowledgments}
\change{We thank an anonymous referee for helpful comments.} We thank Marcus Lower, and Apurba Bera for useful discussions.  \change{We thank Jackson Taylor and Scott Ransom for help with Algorithmic Pulsar Timing.}
RS is supported by NSF grant AST-1816904. DK  is supported by NSF grants AST-1816492 and AST-1816904.  AA is supported by NSF grant AST-1816492.  N.H.-W. is the recipient of an Australian Research Council Future Fellowship (project number FT190100231).
This scientific work uses data obtained from Inyarrimanha Ilgari Bundara / the Murchison Radio-astronomy Observatory. We acknowledge the Wajarri Yamaji People as the Traditional Owners and native title holders of the Observatory site. CSIRO’s ASKAP radio telescope is part of the Australia Telescope National Facility (\url{https://ror.org/05qajvd42}). Operation of ASKAP is funded by the Australian Government with support from the National Collaborative Research Infrastructure Strategy. ASKAP uses the resources of the Pawsey Supercomputing Research Centre. Establishment of ASKAP, Inyarrimanha Ilgari Bundara, the CSIRO Murchison Radio-astronomy Observatory and the Pawsey Supercomputing Research Centre are initiatives of the Australian Government, with support from the Government of Western Australia and the Science and Industry Endowment Fund. 
The Parkes radio telescope is part of the Australia Telescope National Facility which is funded by the Australian Government for operation as a National Facility managed by CSIRO. We acknowledge the Wiradjuri people as the Traditional Owners of the Observatory site.
The Australia Telescope Compact Array is part of the Australia Telescope National Facility (https://ror.org/05qajvd42) which is funded by the Australian Government for operation as a National Facility managed by CSIRO. We acknowledge the Gomeroi people as the Traditional Owners of the Observatory site.
This paper includes archived data obtained through the CSIRO ASKAP Science Data Archive, CASDA (http://data.csiro.au). This research has made use of the NASA/IPAC Infrared Science Archive, which is funded by the National Aeronautics and Space Administration and operated by the California Institute of Technology.  Based on data products from observations made with ESO Telescopes at the La Silla Paranal Observatory under programme ID 177.D-3023, as part of the VST Photometric H$\alpha$ Survey of the Southern Galactic Plane and Bulge (VPHAS+, www.vphas.eu).
\end{acknowledgments}

%

\vspace{5mm}
\facilities{ASKAP, IRSA, Parkes, Swift(XRT), WISE (\url{https://www.ipac.caltech.edu/doi/irsa/10.26131/IRSA153}), ATCA, VST.}


\software{\change{APTB \citep{2023arXiv231010800T}}, astropy \citep{2013A&A...558A..33A,2018AJ....156..123A},  PINT \citep{2021ApJ...911...45L}, PSRCHIVE \citep{2004PASA...21..302H}, PsrPopPy \citep{2014MNRAS.439.2893B}, , Pulsar Survey Scraper \citep{2022ascl.soft10001K}, PyGEDM \citep{2021PASA...38...38P},  reproject \citep{2020ascl.soft11023R}, swarp \citep{2002ASPC..281..228B}.}



\bibliography{references}{}
\bibliographystyle{aasjournal}




\end{document}